# Where's the Structure? A Systematic Literature Review of Empirical Research on Human-AI Collaboration and Hybrid Intelligence for Learning

Luis P. Prieto

GSIC-EMIC research group, Universidad de Valladolid (Spain)

luispablo.prieto@uva.es

Juan I. Asensio-Pérez

GSIC-EMIC research group, Universidad de Valladolid (Spain)

juaase@tel.uva.es

María Jesús Rodríguez-Triana

GSIC-EMIC research group, Universidad de Valladolid (Spain)

mjrodtri@uva.es

Mohamed Saban

GSIC-EMIC research group, Universidad de Valladolid (Spain)

GICAP research group, Department of Digitization, Universidad de Burgos (Spain)

msaban@ubu.es

Yannis Dimitriadis

GSIC-EMIC research group, Universidad de Valladolid (Spain)

yannis@tel.uva.es

**Author Notes**

Luis P. Prieto: https://orcid.org/0000-0002-0057-0682

Juan I. Asensio-Pérez: https://orcid.org/0000-0002-1114-2819

María Jesús Rodríguez-Triana: https://orcid.org/0000-0001-8639-1257

Mohamed Saban: https://orcid.org/0000-0003-3412-8574

Yannis Dimitriadis: https://orcid.org/0000-0001-7275-2242

Contact information: Luis P. Prieto, School of Telecommunications Engineering, Paseo Belén 15. Campus Miguel Delibes. 47011 Valladolid, Spain. Email: luispablo.prieto@uva.es

**Abstract**

Artificial intelligence (AI) has been applied across educational contexts to support learning. One approach to such support is "human-AI collaboration" (also termed "hybrid intelligence"), where human(s) and AI components interact to promote human learning. However, as in human-to-human computer-supported collaborative learning (CSCL), unstructured interaction does not necessarily produce an effective learning experience. This paper reports a systematic literature review of empirical studies (N=62) on human-AI collaboration and hybrid intelligence for learning support. The review characterizes collaboration processes, their structures, and contexts of application. It also extracts emerging design knowledge and research gaps. Researchers and technology designers can use these findings as a starting point for structuring more effective AI-enhanced technologies for collaboration, in educational practice and future research.

*Keywords:* Human-AI collaboration, Artificial intelligence in education, Literature reviews, Hybrid intelligence, Computer-supported collaborative learning.

## Introduction

Collaborative learning (CL) is often defined as "a situation in which two or more people learn or attempt to learn something together" (Dillenbourg, 1999, p. 1). CL has been characterized in multiple ways (e.g., equality and mutuality of influence, group processing, positive interdependence, ... see O'Donnell & Hmelo-Silver, 2013), with no universally accepted definition. Nevertheless, decades of research in (computer-supported) CL have evidenced its many key benefits: improved social outcomes, higher achievement, more time on task, better psychological health, higher-quality reasoning, etc. (Johnson & Johnson, 2009; O'Donnell & Hmelo-Silver, 2013).

Research on collaborative and cooperative learning has also shown that these benefits vary depending on the *structure* of learner interactions (e.g., Slavin, 1996). Many researchers have observed that "the richness of social interactions (conflict resolution, elaborated explanations, mutual regulation, ...)" predicts learning outcomes, prompting the design of scripts for "enhancing the probability that these knowledge productive interactions occur during collaboration" (Dillenbourg & Jermann, 2007, p. 276). A significant strand of (CS)CL research has thus investigated scripted collaboration (Kollar et al., 2006) and proposed recurring patterns of effective CL, such as *Jigsaw*, *Pyramid* (see Hernández Leo et al., 2005), or *ArgueGraph* (Jermann & Dillenbourg, 2003).

A common distinction in this literature is between *macro-* and *micro-*scripting (Kobbe et al., 2007). Macro-scripts describe activities over longer time segments across multiple social planes (individual, small-group, class-wide), specifying both what learners do and how (Dillenbourg & Jermann, 2007). Micro-scripts, in contrast, provide finer-grained, psychologically-oriented scaffolding to individual students, such as sentence starters and question prompts (Kollar et al., 2006).

Against this backdrop, recent advances in artificial intelligence (especially, generative AI) have prompted the emergence of "human-AI collaboration" (HAIC) as a potential strategy to support human learning. HAIC (sometimes called "human-AI teaming") has a longer history in human-computer interaction, though a widely accepted definition remains elusive. Fragiadakis and colleagues (2025) describe it as "a symbiotic partnership where humans and AI work together to achieve shared goals", typically grounded in the complementary strengths of human and artificial intelligence (e.g., intuitive vs. analytic reasoning; Dellermann et al., 2019). A related concept, "hybrid intelligence" (HI), draws on distributed cognition (Hutchins, 1995) and frames the combination of human and artificial intelligences as a single, unified system, defined as "the ability to accomplish complex goals by combining human and artificial intelligence to collectively achieve superior results than each of them could have done in separation and continuously improve by learning from each other" (Dellermann et al., 2021).

With recent advances in machine learning and generative AI (GenAI), AI systems have gained considerable flexibility in human interaction and increasingly agentic features (see reviews by Mittal et al., 2024; Noroozi et al., 2024). As a consequence, HAIC and HI have been proposed as useful conceptualizations to specifically support human learning (Cukurova, 2025; Gyasi et al., 2025; Järvelä et al., 2025).

Given what we know about the variable effectiveness of unstructured human collaboration, it stands to reason that HAIC and HI will face similar problems without structures designed to foster productive human-AI interactions. This assumption appears increasingly validated by studies documenting the deleterious effects of learners' GenAI use, from overreliance (Hou et al., 2025) to what some researchers call "metacognitive laziness" (Y. Fan et al., 2025).

This paper examines the current empirical evidence on the use of HAIC and HI for human learning, focusing specifically on the HAI/HI collaboration *process*: the "how" of such collaboration (Labadze et al., 2023). Our aim is to *understand what kinds of HAI/HI collaboration structures are being investigated by researchers*, with the ultimate goal of producing design knowledge (e.g., HAIC/HI "design patterns" for learning, cf. Alexander et al., 1977) reusable by researchers, educational technology designers, and practitioners. This also relates to key concerns in the AI in education (AIED) field, such as tensions between human control vs. AI automation (see, e.g., Cukurova, 2025; Holstein et al., 2020; Molenaar, 2022), and learner agency (Cukurova, 2026; Darvishi et al., 2024).

To this end, we conducted a systematic literature review (SLR) of empirical studies on HAIC and HI *to support learning* (vs. AI uses that support or automate teaching, see e.g., Dutta & Kannan Poyil, 2025 in workplace learning; or Rokkones & Giannakos, 2025 in K-12). We describe the methods used to narrow 1,632 retrieved sources to an in-depth analysis of N=62 papers, present our findings, and discuss implications, emergent design knowledge, limitations, and future research lines.

## Related Work: Literature Reviews in Human-AI Collaboration and Hybrid Intelligence for Learning

To motivate this systematic literature review, we situate our work among prior reviews in adjacent areas. Reviews of HAIC exist, but they mostly follow an HCI tradition and emphasize how HAIC supports (often business) tasks in organizations (M. Braun et al., 2023), or in decision support systems (Gomez et al., 2025). Reviews of HI show a similar bias: Wiethof and Bittner (2021) examine how humans and AIs learn together in HAIC and extracts recurring HAI "collaborative learning processes" (e.g., decision support by explanation, exploration, etc.). Hemmer et al. (2021) focus on collaboration in explainable AI (XAI) for decision support systems. Kamar (2016) centers on how humans help AIs learn (the opposite of our aim here), and Pileggi (2024) focuses on particular technologies (ontologies) in HI. Despite their merits, all these reviews have aims and concerns remote from educational technologies. Our aim is thus closest to Wiethof's and Gomez's reviews in our searching for interaction structures and collaboration processes, but with the explicit goal of *supporting human learning*.

High-level overviews of AIED also exist. Beyond reviews already mentioned (like Mittal et al., 2024; Noroozi et al., 2024), Zawacki-Richter et al.'s (2019) classic review identifies four key application areas of AI in education, yet it does not cover HAIC in detail. Chiu (2024) does include a "learning" domain (i.e., AI to enhance learning) and identifies "roles of AI", including "human-machine conversations" and "analyzing student work for feedback". It does not, however, unpack collaboration processes, and its publication date predates the emergence of GenAI. Wang et al.'s (2024) review is more recent, but treats CL support as a small sub-category (5 studies) and largely represents a flavor of HAIC where the AI facilitates or monitors human–human interaction. Yang et al. (2025) review GenAI applications for CL, but mainly characterize form factors (e.g., chatbots), kinds of learning (e.g., self-regulated learning), learning outcomes, and broad CL aspects (e.g., argumentation), without analyzing HAIC/HI processes and their structures.

Within secondary research on AI in education, some reviews target specific system types. Intelligent tutoring systems (ITSs) are well covered (e.g., Du Boulay & Luckin, 2016), but we contend they primarily substitute teaching tasks (e.g., through exercise delivery towards an adaptive learning pathway) rather than directly interacting or learning together with the learner; without interaction, there cannot be collaboration (cf. Dillenbourg, 1999). Reviews of educational chatbots are actually closer to HAIC for learning. Two literature reviews (Okonkwo & Ade-Ibijola, 2021; Wollny et al., 2021) synthesize benefits and challenges of such systems, but not the *process* of collaboration; they also omit non-chatbot HAIC, and also cover non-learning uses (e.g., teaching support, frequently asked questions), predating the recent GenAI wave. More recent chatbot reviews (Deng & Yu, 2023; Labadze et al., 2023) likewise emphasize benefits and disadvantages rather than *how* such benefits are achieved, and again exclude other AI system formats that could foster HAIC/HI for student learning. Deng and Yu do synthesize roles chatbots play (e.g., "learning partner," perhaps "teaching assistant"), which connects to the collaboration structures we seek. Han and colleagues' (2025) scoping review of intelligent "learning companion systems" (LCS) overlaps more closely our HAIC/HI for learning focus: LCSs simulate human-like characteristics to provide support and companionship. Han and colleagues enumerate up to 14 AI roles, from "motivator" or "goal setter" to "co-learner" or "troublemaker." Although narrower, this role taxonomy does offer actionable design knowledge, analogous to roles in (human-to-human) CSCL scripting.

Finally, reviews of AI for collaborative learning (which, by definition, includes HAIC/HI for learning) offer relevant coverage but remain AI-centric. Tan et al. (2022) categorize techniques (e.g., clustering, natural language processing – NLP) and goals, but does not detail pedagogical mechanisms, processes, or structures, and it predates the GenAI surge. Kovari (2025), on the other hand, reviews AI applications in higher education (HE) and foregrounds algorithms, including CL supports that are not necessarily HAIC/HI (e.g., recommender systems or personalizers). While they stress "good task design", they provide limited guidance on what interaction structures constitute such good design.

We therefore conclude that, to the best of our knowledge, no existing literature review specifically targets *HAIC and HI used for student learning*, analyzes the HAI/HI collaboration/interaction *process* and its structures in detail, and is recent enough to include the recent wave of *GenAI-based applications*. This gap motivates the review presented in the remainder of this paper.

## Materials and Methods

### Research Questions

To understand the current state of empirical research on HAIC and HI for learning, and the collaborative processes and structures researchers have proposed, we conducted a systematic literature review following PRISMA guidelines (Page et al., 2021). This SLR adopted a mixed-methods approach, emphasizing qualitative and interpretive analysis over quantitative synthesis, given the heterogeneous and exploratory current state of this research area. The SLR addressed the following research questions, each illuminated through multiple informative questions (IQs) in an anticipated data reduction scheme (see Fig.1) (Miles & Huberman, 1994):

- RQ1: What kinds of empirical research have been conducted about HAIC/HI for (human) learning?
- RQ2: How is HAIC/HI conceptualized in this body of work?
- RQ3: How are HAIC/HI (for learning) processes structured in this body of work?
- RQ4: What design knowledge and research gaps can be derived from this body of research?

**Figure 1**

*Overarching research question, particular research questions, and informative questions explored in the systematic literature review*

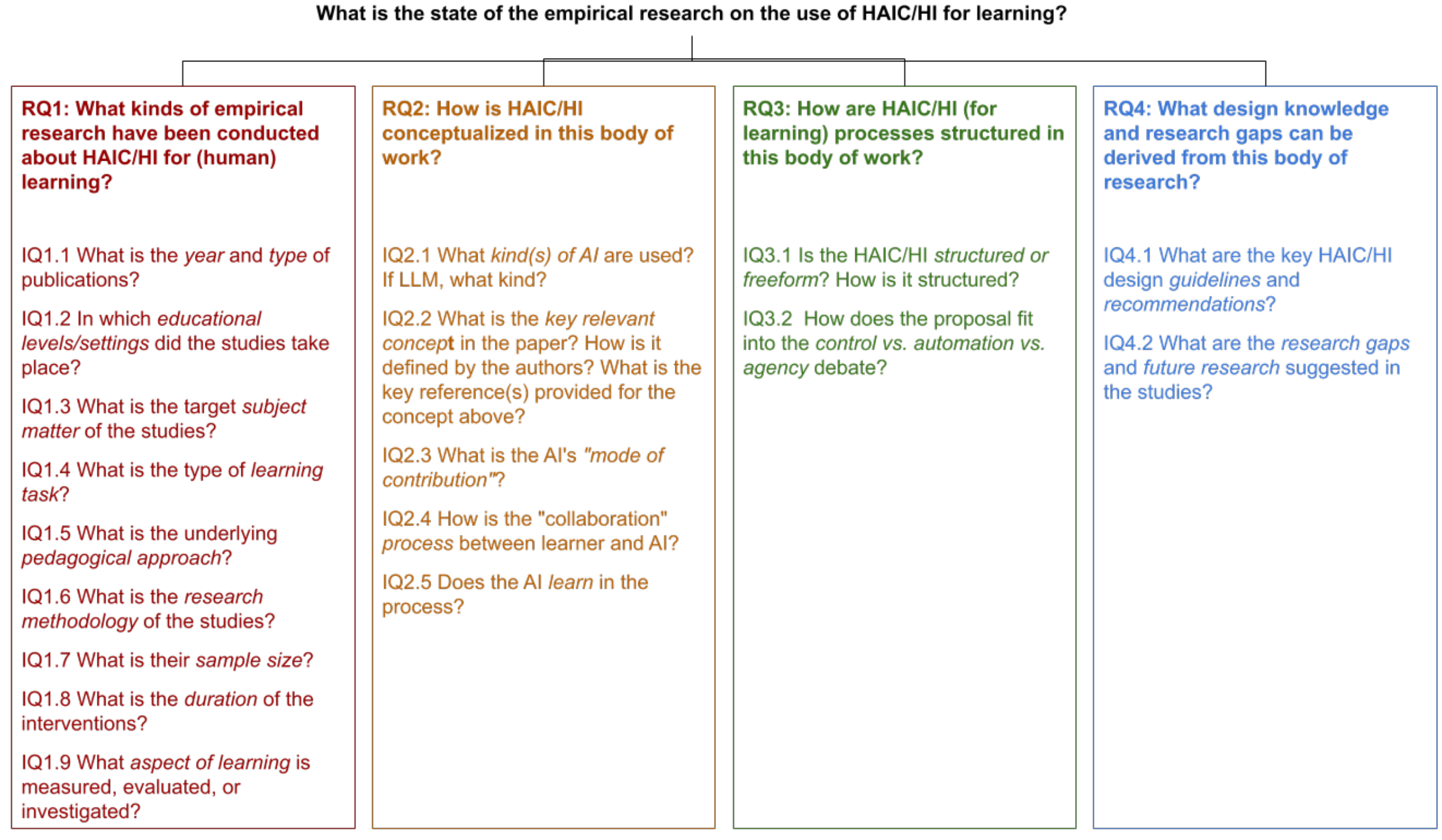

## Review Method

*Eligibility criteria*. We sought papers reporting empirical studies (i.e., reporting both methods and nontrivial empirical data) in the educational domain, focused on human learners interacting with AI elements toward the goal of learning. *Exclusion* criteria were:

- Not an individual article (e.g., a full conference proceedings volume);
- Full text not retrievable;
- No abstract available;
- Not written in a language accessible to the research team (primarily English and Spanish);

- Not aimed at the educational domain;
- Not an AI empirical study;
- The AI's objective is not to improve student learning (e.g., focuses on AI-teacher collaboration, or uses a student population but not paying attention to a learning process);
- Learners do not directly interact with an actual AI (e.g., Wizard-of-Oz studies in which the student interacts with a human posing as an AI, student surveys about past independent use of AI, or expert panels); and
- No details provided about the HAIC/HI process, or HAIC/HI not mentioned significantly in the text of the paper.

*Information sources*. Given the current state of flux of all AI-related fields and the large amounts of unvetted grey literature (e.g., preprints) being produced, and the fact that we try to encode best practices in HAIC/HI for learning, we prioritized established, peer-reviewed sources over works-in-progress and grey literature. We searched the two best-established multidisciplinary databases relevant to educational technologies: Web of Science (WoS)[1] and SCOPUS[2]. These databases subsume additional field-specific databases such as ACM and IEEE.

*Search strategy*. After an initial scoping of common synonyms beyond HAIC and HI (e.g., "human-AI teaming"), we built a query to capture the overlap between the HAIC/HI *concept* and the *education/learning application* research space. Other exclusion criteria were checked later, based on the paper's metadata and full text. The query was applied to the titles, abstracts, and keywords of the sources:

---

[1] https://www.webofscience.com/
[2] https://www.scopus.com/

*("human-AI team" OR "human-artificial intelligence team" OR "human-AI collaboration" OR "human-artificial intelligence collaboration" OR "hybrid human-AI" OR "hybrid human-artificial intelligence" OR "hybrid human-system" OR "hybrid intelligence" OR "AI assistant" OR "artificial intelligence assistant")* AND *("teaching" OR "tutoring" OR "coaching" OR "learning" OR "teacher" OR "educator" OR "instructor" OR "learner" OR "student" OR "education" OR "educative").*

We ran a first retrieval on May 2nd, 2024. Because research accelerated after the widespread use of GenAI technologies such as large language models (LLMs), we ran a second, updating retrieval on Apr 30th, 2025, restricted to years 2024-2025.

*Study selection process*. As shown in Figure 2[3], the first wave returned 869 records (640 SCOPUS; 229 WoS) and the second returned 763 (572 SCOPUS; 191 WoS). After deduplication, five researchers applied inclusion/exclusion criteria in two stages. In Stage 1, the team screened titles, keywords, and abstracts, keeping unclear cases for the next phase. In Stage 2, full texts were screened to confirm empirical methods and data were present, and to verify that the HAIC/HI process was described in sufficient detail to code. In both stages, we began with all team members screening a shared sub-sample; disagreements and doubts were discussed until we reached full consensus on that sub-sample. We then distributed the remaining records across the team for independent screening. At the end of this process, the final corpus comprised N=62 papers for in-depth analysis aligned with the RQs.

**Figure 2**

*Source search and filtering process (cf. PRISMA guidelines)*

[3] See more details in the additional materials at https://osf.io/4wxh2/overview?view_only=9fcbf8ed4bdb405b8d812986f2c1c966

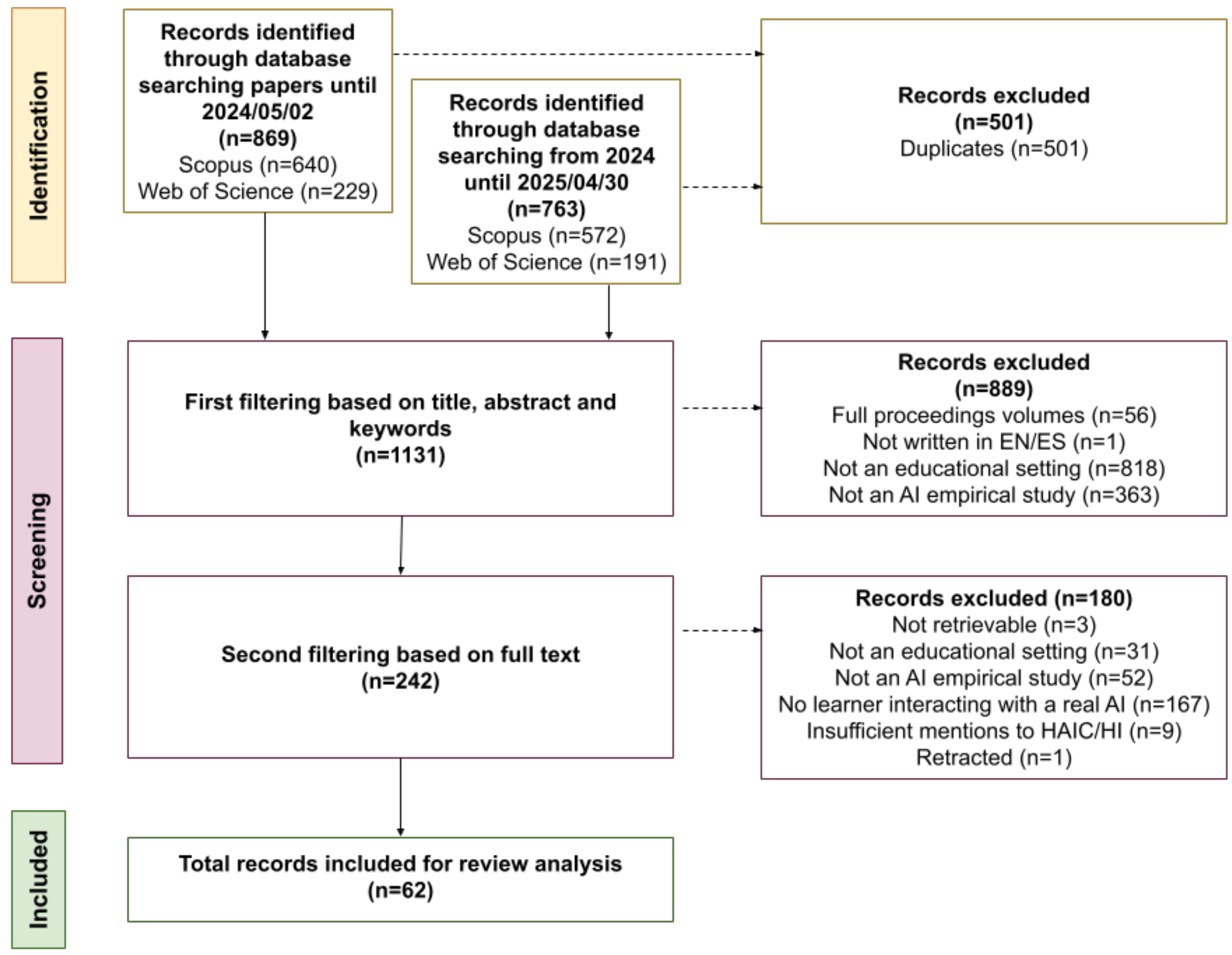

*Data collection process*. Full-text sources were analyzed manually using Google Forms/Spreadsheets, via a structured form corresponding to the IQs in Figure 1. The form combined closed categories (e.g., educational level in IQ1.2; micro-/macro-type of collaboration structure as part of IQ3.1) with open responses (e.g., how is the collaboration process between learner and AI? IQ2.4). All team members independently coded a purposefully selected subset of four sources; disagreements and doubts were discussed until resolved, sharpening shared understanding of the questions and categories. The remaining sources were then distributed among the team members.

*Synthesis approach*. Closed-response IQs were synthesized using descriptive and visual statistics. For the majority of IQs, captured through open-response questions, we employed a qualitative synthesis approach using an inductive process modeled on reflexive thematic analysis (V. Braun & Clarke, 2006). This approach treats themes as researcher-constructed interpretations rather than fixed, discoverable objects. For each IQ, one researcher reviewed all form responses (thus ensuring that each paper-IQ pair was examined by at least two researchers), disagreements were resolved, and the resulting categories and interpretations were discussed collectively by the whole team.

# Results

## RQ1: The State of Empirical Research in HAIC and HI for Learning

Looking at basic descriptors of the analyzed papers (see also Figure 3 below), we observe a sharp recent increase in research in this area (year of publication, IQ1.1 in Figure 1), coinciding with the public emergence of GenAI. In fact, 2024 alone yielded more studies meeting our criteria (empirical works using HAIC/HI to support learning) than all previous years combined (Figure 3, top-left). The predominance of journal articles over conference papers (37 vs. 23) largely reflects our aim of identifying more established work (and the databases selected).

Regarding educational settings (IQ1.2, Figure 3, top-right), most studies were conducted in higher education (57/62 studies). Only a few addressed other levels, such as C.-H. Lin et al.'s (2025) exploration of GenAI for second-language writing in K-12, or were level-agnostic, such as Chen et al.'s (2024) study of how an LLM chatbot is used in modeling tasks with NetLogo. This imbalance is typical of early-stage fields, in which academics (mostly working in higher education) experiment with technologies in readily available settings. It may also reflect ethical concerns about testing experimental technologies with underage learners.

The subject matter of the supported learning activities (IQ1.3, Figure 3, bottom-left) was quite varied. A large minority of studies focused on information technology and computer science (19/62), which is unsurprising given the likely disciplinary origins of many researchers in the area. There was also notable representation of language, literature, and writing support (9/62), including Song et al.'s (2022) use of a pre-GenAI voice assistant to support second-language learner pronunciation skills, or Taeza's (2025) use of LLMs for writing practice. Design and other creative disciplines were also prominent (11/62, e.g., Gual-Ortí et al., 2025). Although these areas are not always central in other educational technology innovations, they align well with GenAI's fluency in producing text and images, common outputs in such fields. They also fit HAIC/HI pedagogically, as these disciplines are often taught through teamwork. The learning tasks supported by HAIC/HI (IQ1.4) showed a similar pattern: writing tasks and design projects were especially common (13 papers each), followed by computer programming (7 papers, e.g., Hussein et al., 2024), alongside a long tail of activities ranging from questionnaires and reflection to embodied exercises. Together, these suggest the flexibility of AI, especially GenAI, technologies.

From a pedagogical perspective (IQ1.5, Figure 3, bottom-right), the large number of studies that did not define a clear pedagogical approach (22 papers) again suggests an early-stage, technology-driven research area. Among studies that did specify a pedagogy, the most common were collaborative learning (18 papers, e.g., Ding & Chiu, 2024) and project-based learning (14 papers), which is consistent with our focus on HAIC/HI and with the fact that projects are often developed through teamwork.

**Figure 3**

*Descriptive characteristics of the N=62 reviewed papers.*

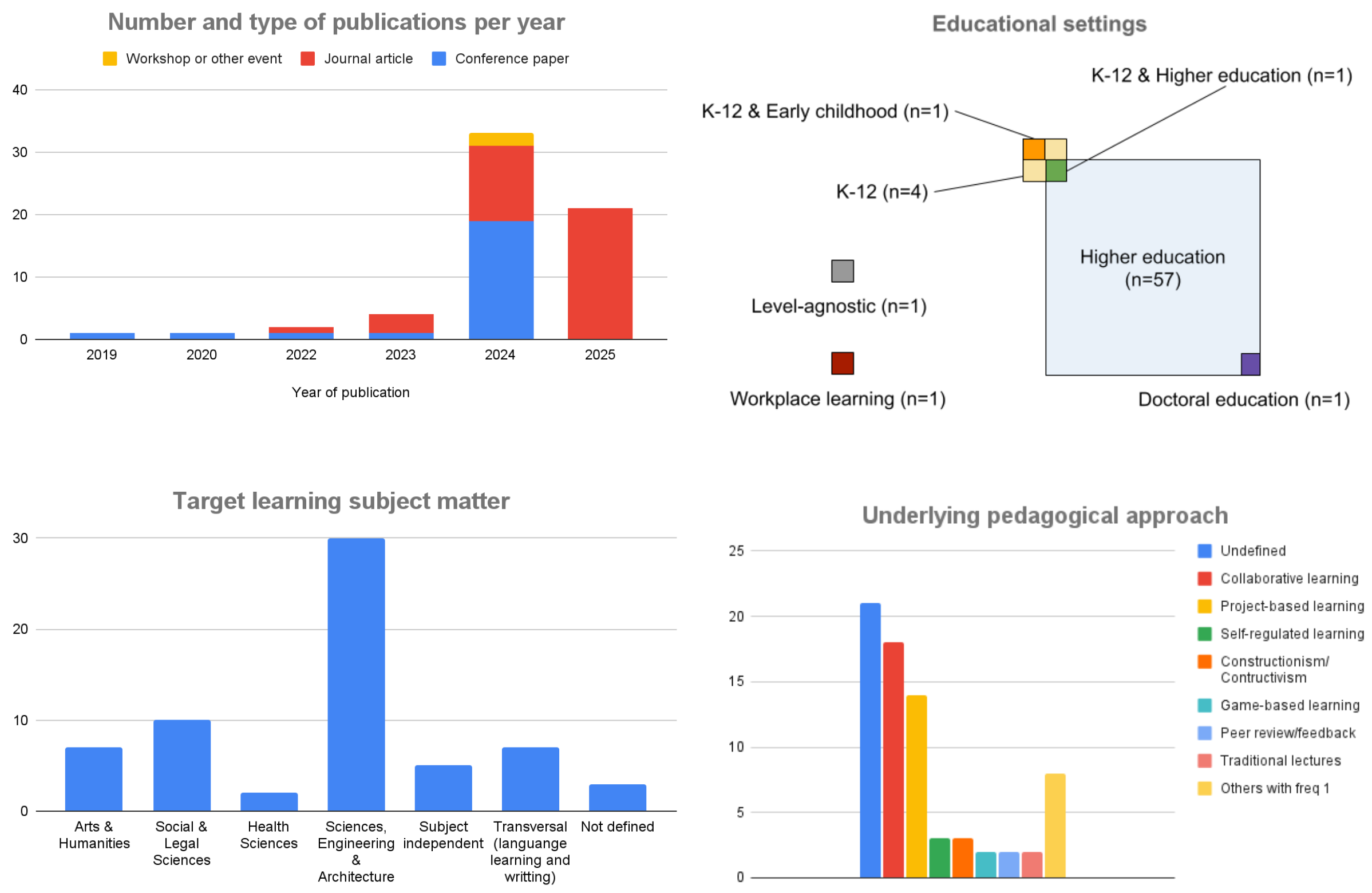

The reviewed papers' research designs (IQ1.6) also reflect an early-stage, technology-driven field. Most were case studies exploring technological possibilities (28 papers), followed by experimental studies comparing implementations (18 papers), for example "with versus without" AI support or across different AI conditions, such as Kumar et al.'s (2024) comparison of LLM-based reflections against no-reflection or lecture controls. We also identified 8 survey-based studies (e.g., Rasnayaka et al., 2024), along with several other less common methodologies. Sample sizes (IQ1.7) were likewise heterogeneous. Most studies involved tens of participants (38 papers), while 13 were conducted at the scale of hundreds, pointing to the scalability potential of this technology even at an early stage. Notable examples include Darvishi et al. (2024) and Hou et al. (2025) (with 1,625 and 1,818 participants, respectively), which used AI to support peer review of learning resources at scale and to develop a questionnaire instrument measuring modes of student reliance on AI (an aim that inherently requires large samples).

The duration of the HAIC/HI interventions studied—that is, the period over which learners and AI systems interacted (IQ1.8)—was also highly variable, both in researchers' reporting and in actual length. Among studies providing enough detail on number and length of sessions and session design, many used single-session formats, such as Cai et al.'s (2024) study of a chatbot participating in small-group learner conversations for 20 minutes in one session. This limitation has also been noted in recent critiques of GenAI-for-education studies (Weidlich et al., 2025). We did, however, find longer deployments lasting multiple weeks or a semester (e.g., Parker et al., 2025). A particularly notable case is G. Fan et al. (2025), in which LLM support in programming courses was deployed over two academic years. At the same time, the longer interventions often appeared to involve limited structure or little in-depth observation of learner-AI interactions.

Finally, regarding the aspects of learning examined in the analyzed papers (IQ1.9), target measures were diverse. Most studies (35/62) examined some aspect of the learning process, from qualitative analyses of conversations or video recordings to quantitative eye-tracking measures. General user experience measures, such as technology acceptance or suggestions for improvement, were also common (32 papers), as were measures of product/outcome quality (27 papers, e.g., Kim et al., 2024). By contrast, only 13 studies measured learning gains, even though this might be expected to be the central outcome of educational technology interventions. This pattern echoes recent critiques of the area (Yan, Greiff, et al., 2025; Yan, Pammer-Schindler, et al., 2025).

**RQ2: Characterizing HAIC and HI for Learning**

To understand the kinds of HAIC/HI being empirically studied, we first examined the types of AI used in interventions (IQ2.1). Generative AI was the most studied (45/62), predominantly LLM-based systems (e.g., ChatGPT), with a few studies using image-generating GenAI (e.g., Stable Diffusion, Midjourney - see Hwang et al., 2025; or Liu et al., 2025). Eight studies featured classic machine-learning systems and another eight used non-LLM natural language processing. In rare cases, multiple AI types were combined (e.g., Weber et al., 2025 use of traditional machine learning with BERT to provide structured feedback on legal writing). Among LLM-based studies, 43/45 relied on online services (most commonly OpenAI's ChatGPT app or API), with only four using open-weight, locally-executed models (e.g., Weber et al., 2025 using BERT) or remaining model-agnostic (a framework that can use different models as backend, in Alier et al., 2025).

Regarding the key concept of interest in the reviewed papers (IQ2.2), HAIC was the most frequently used concept (45/62) versus 9 papers centered on HI; 17 papers revolved around "AI assistants". Surprisingly few studies explicitly defined HAIC. The six HAIC definitions identified (plus two via snowball sampling) varied considerably, but most denoted a *task being done together*, emphasized *complementarity of human and AI abilities*, and in some cases stressed *augmenting* rather than replacing humans while keeping *responsibility and control with humans*. Some definitions focused on *task-oriented goals* (e.g., improving performance), others on *person-oriented goals* (e.g., enhancing cognition). For instance, Gyasi et al. (2025) propose that "AI and humans work together by taking advantage of the capabilities of each party to boost performance". Among the 9 HI-centered studies, definitions emphasized cognition more heavily (the combination of human and machine intelligence) though *complementarity* and *task/person* *goals* remained similarly prominent. Y. Fan et al. (2025) (citing Akata et al., 2020) define HI as the "combination of human and machine intelligence, augmenting human intellect and capabilities instead of replacing them and achieving goals that were unreachable by either humans or machines" (p. 491). Notably, none of the reviewed studies reconceptualized these general definitions for learning-specific contexts. The work of Dellermann and colleagues (2021) defined HI as "the ability to accomplish complex goals by combining human and artificial intelligence to collectively achieve superior results and continuously improve by learning from each other." This definition appeared in both HAIC and HI studies (2 and 3 times, respectively), suggesting conceptual interchangeability of those two concepts among researchers. The most-cited education-specific definitional work is Järvelä et al. (2023) (itself pointing to Akata et al., 2020), alongside Holstein et al. (2020), who use "hybrid adaptivity" to describe humans and AIED systems complementing each other toward improved task performance and/or co-learning. These references and definitions are consistent with similar works oriented towards teaching (e.g., Rokkones & Giannakos, 2025).

Analyzing the AI's mode of contribution in the learner-AI team (IQ2.3, see Cui & Yasseri, 2024) further triangulates researchers' conceptualizations. A large majority (52/62 papers) used an "AI assistant" mode, in which humans offload tasks or expect answers from the AI (e.g., a ChatGPT chatbot). Despite the HAIC/HI framing, only 10 studies portrayed a "teammate AI" — in which the AI operates as a roughly equal team member (e.g., a system promoting reflection after video analysis, in E. Yu, 2023). Twelve works used a "coach AI" mode, providing feedback (e.g., the agent-in-the-loop supporting reflection in collaborative programming, in Sankaranarayanan et al., 2020). Only one work used an "embodied AI" mode or explicitly supported multiple modes (the framework for educational agents in learning management systems by Alier et al., 2025). Notably, several reviewed works exhibited "mislabeling" or "metaphor stretch". For example, the introduction of vanilla AI chatbots to which students offloaded tasks was presented as "AI introduced as a co-designer" (Adorni & Grosso, 2025, p. 1) or as a teammate (Darban, 2024), reflecting the terminological fuzziness typical of early-stage research areas (cf. Borup et al., 2006; Rip & Voß, 2013).

An inductive analysis of collaboration setups (IQ2.4) showed that most studies gave learners a *vanilla, non-customized AI chatbot* for student-initiated "collaboration" (34 studies), with some providing a *pedagogically-modified chatbot* for similar assistant-like cooperation (15 studies; e.g., CodeAid, which offers specialized functions to support coding learning, in Kazemitabaar et al., 2024). In 18 studies, learner-AI interaction was mediated by a *socially-enforced scaffold* (typically, written or oral teacher instructions) which could be general (e.g., prompting tips, as in Garg et al., 2025; or Goel et al., 2023) or process-specific (e.g., designated AI use phases, as in Min et al., 2025). Technological scaffolding integrated into system interfaces was less common (6 papers). Another recurring setup inserted the *AI as an actor in human-to-human conversations* (6 papers; e.g., Gutiérrez-Ferré et al., 2024). While learner initiative dominated, notable exceptions exist: In Chen et al. (2024), the AI actively gathered information to adapt future interactions. Less frequent setups included embodied or virtual reality simulations (3 papers, e.g., Tracy & Spantidi, 2025), HCI-focused assistants (Da S. Soares et al., 2019), and humans collaborating to build AI models (Mlynář et al., 2024). Perhaps most unexpectedly, genuine *co-learning processes*, in which humans and AI share the same task or goal, were rare. Bosch et al.'s (2025) learning-oriented human-AI teaming research (where humans and simulated robots collaboratively learn about a rescue situation) and, to a lesser extent, Yu et al. (2025), are notable exceptions.

This connects to whether AI "learns" through its interaction with human learners, that is, whether any model update occurs during collaboration (IQ2.5). Aside from Bosch et al. (2025), the answer is an overwhelming *no*. Some LLM-based systems retain past interactions in context to adapt responses (e.g., Chang et al., 2024), but that falls well short of reinforcement-learning-style model updates, which are known to have both benefits and dangers (Whittlestone et al., 2021). The near-total absence of discussion about AI learning across the reviewed papers suggests that researchers treat HAIC/HI more as a metaphor than as a concept to be taken to its full conceptual consequence.

**RQ3: HAIC and HI Structures for Learning**

Turning to our main aim (whether and how HAIC/HI processes were structured, IQ3.1), 24 of 62 papers imposed no structure on human-AI interaction (what we term freeform interaction, e.g., Nguyen, Hong, et al., 2024; Zou & Huang, 2024). The remaining 38 papers did impose some kind of *structure* or constraint. These were fairly evenly split between "*macro* scripts/structures" (18/38) – coarse-grained pedagogical designs that create learning situations within which the desired student interactions occur (cf. Dillenbourg & Tchounikine, 2007) – and an overlapping set of "*micro* scripts/structures" (25/38), which scaffold the interaction itself through devices such as sentence starters or question prompts (cf. Kobbe et al., 2007). As noted under IQ2.4, these structures were often delivered *socially* (15/38, e.g., teacher instructions as in Garro Mena, 2024) but were sometimes also *embedded in technology* (15/38, e.g., user interface phases/transitions, as in Weber et al., 2025). In 21 of those 38 studies, the constraints were additionally driven by LLM prompting (as in CodeTutor, Lyu et al., 2024) or pre-training of BERT-derived models, carried out by teachers and/or researchers.

Beyond the micro/macro distinction, our inductive analysis of the HAIC/HI interactions reported in the reviewed studies identifies three recurrent, sometimes overlapping collaboration structures, sometimes purposefully imposed by the system, teachers, or researchers – but sometimes emerging spontaneously (see Figure 4):

1. *"AI creates/answers, humans refine" (46/62 papers)*: Students delegate tasks or pose questions to the AI, then assess the AI-generated artefacts or answers, deciding whether further refinement is needed and engaging in multiple cycles of human feedback and AI refinement, either by modifying outputs themselves or by issuing additional requests to the AI. In three of the studies, students make iteratively refined requests to the AI (all three cases involving image generation) without making their own modifications to what the AI generates (Cao et al., 2025; Han et al., 2024; P.-Y. Lin et al., 2024). Regarding the cyclic nature of these interactions, Hou et al. (2025) is the sole case in which students request AI assistance only once without subsequent engagement. In all other papers using this structure, students actively refine AI-generated content across multiple human-AI cycles of engagement. This structure maps most directly to the "assistant" mode of AI contribution described above (see IQ2.3).
2. *"Humans create, AI assesses" (18/62 papers)*: Students carry out a learning task, typically producing an artefact such as a document or drawing, and at various points may request AI assessment and feedback. That feedback may be structured by a pedagogical framework (akin to a micro structure) or constrained loosely, or not at all, e.g., through LLM prompting. Students then decide how to incorporate it and may engage in further cycles of human refinement and AI feedback. Feedback types include direct performance assessment (e.g., Cosentino et al., 2025; Song et al., 2022), suggestions for improving student proposals (e.g., Y. Fan et al., 2025; Min et al., 2025), and scaffolding for reflection (e.g., Kumar, Xiao, et al., 2024; Parker et al., 2025). Aslan et al. (2024) and Cosentino et al. (2025) are the only two papers in which students receive AI feedback just once, without engaging in additional interaction cycles. This structure maps to the "coach" mode of AI contribution (see IQ2.3).

3. *"AI participation in human collaboration" (7/62 papers)*: Students collaborate through computer support (e.g., a chat), and the AI monitors that collaboration, intervening in three ways: (a) providing an assessment of the ongoing human collaboration (Cai et al., 2024; Sankaranarayanan et al., 2020) – for example, a chatbot detecting unequal participation and suggesting more opportunities for less-active students (Cai et al., 2024); (b) answering a question raised by a participating student (Cai et al., 2024), potentially triggering a shift toward structure #2; or (c) playing the role of another team member (5 papers) – for instance, an LLM-based chatbot prompted to act as a student in a group discussion forming part of a pyramid or snowball collaborative learning macro structure (Gutiérrez-Ferré et al., 2024). Depending on the nature of AI's interjections, this structure relates to either the "coach" or "teammate" modes of AI contribution (see IQ2.3).

**Figure 4**

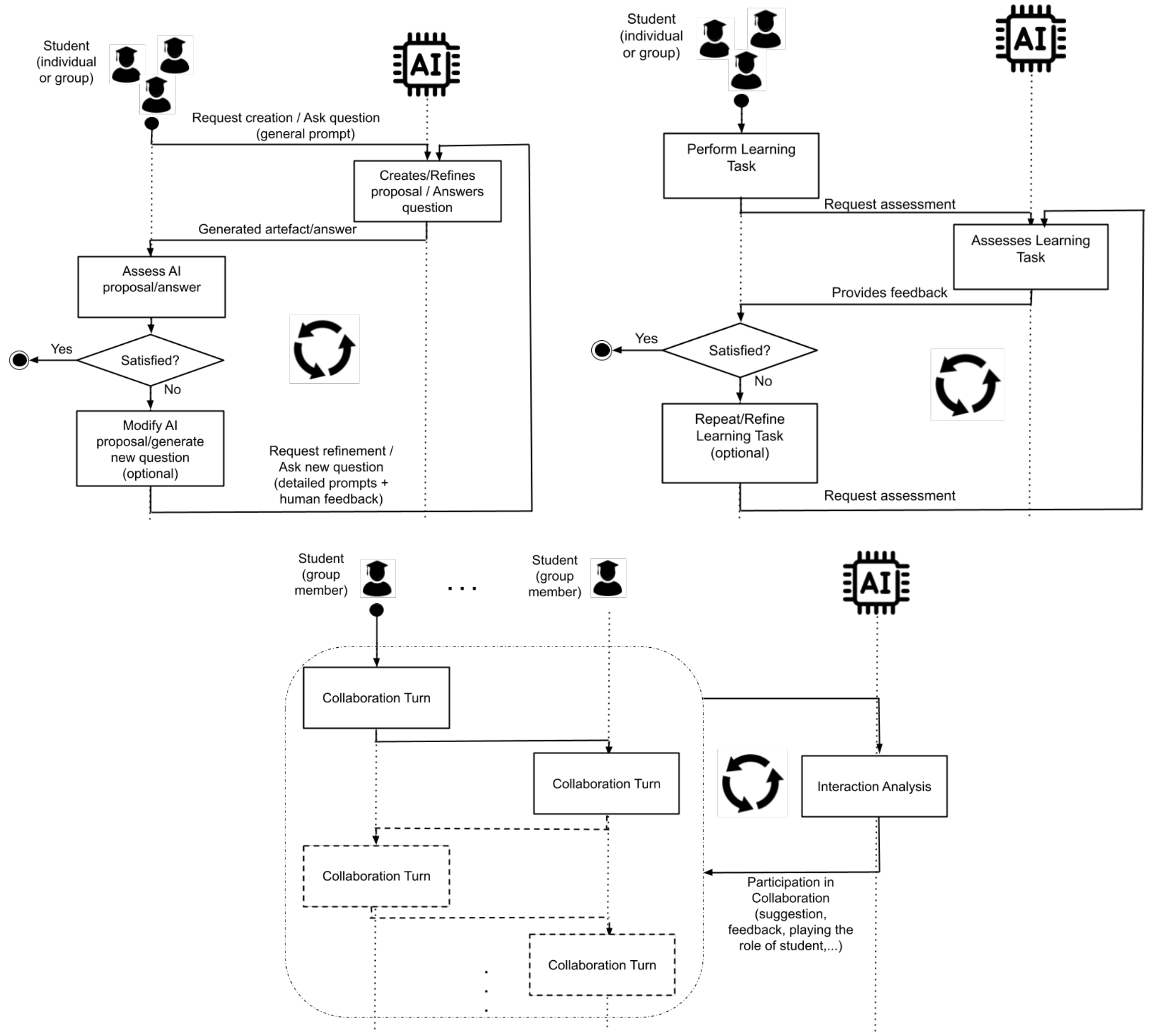

*Basic collaboration structures emerging from the analysis of N=62 HAIC/HI for learning studies: 1) AI creates, humans refine (top-left); 2) Human creates, AI assesses (top-right); 3) AI participation in human collaboration (bottom).*

Beyond these collaboration structures, a key concern in recent literature is the tension among human control vs. AI automation, and agency (IQ3.2) which has implications for the reliability, safety, and trustworthiness of human-centered AI (HCAI) approaches (Shneiderman, 2020). These tensions are well-recognized in AIED (see, e.g., Cukurova, 2025; Darvishi et al., 2024; Holstein et al., 2020; Molenaar, 2022 for recent examples) since certain distributions of control and automation between humans and AI can produce significant learning problems such as over-reliance (Zhai et al., 2024) or "metacognitive laziness" (Y. Fan et al., 2025). Regarding human vs. AI *control*, only in Mlynář (2024) do students exercise full control over the AI tool, as they are the ones actually building the machine learning model. Gyasi (2025) presents another notable case, allowing students to change the AI's "mode of contribution" (see IQ2.3) by switching between "teammate" and "assistant". In all other papers, students either exert no control at all over the AI tool (21/62) or can steer its pre-configured behavior, typically through some kind of prompting (39/62). As our analysis shows, pedagogical control over the AI's behavior (through system constraints and/or prompts) is typically exercised by teachers or researchers rather than by students. Consistent with this low level of human control, AI *automation* (the extent to which the AI carries out its tasks without human intervention) is high in 36 of 62 papers. Three situations account for most of these cases: (1) "Human creates, AI assesses" instances where the AI delivers feedback without additional human input (15 studies; e.g., Darvishi et al., 2024 for peer feedback; Y. Zhang et al., 2026 for writing tasks); (2) "AI creates, humans refine" instances where students do not help refine the AI's initial artefact (14 papers; e.g., Akçapınar & Sidan, 2024 for software development; Goel et al., 2023 for persona creation); and (3) all seven "AI participation in human collaboration" studies, where the AI monitors human-human interaction without receiving additional instructions during the activity. In 25 other papers, automation is lower because the AI requires ongoing human input (typically, additional prompts) to refine and complete its task. Aslan et al. (2024) stands out as a case of very low automation, with the AI apparently controlled and monitored continuously by an instructor.

Finally, student *agency* in HAIC/HI, defined as "the capacity for students to actively shape their learning experiences, make responsible decisions, and control their educational journey" (Darvishi et al., 2024, p. 2), is explicitly examined in only one of the 62 reviewed studies (Darvishi et al., 2024). In 56 of 62 papers, however, students appear to retain a meaningful degree of agency, having the final say on how to act on AI-generated feedback, artefacts, or suggestions. Two exceptions stand out: in Gual-Ortí et al. (2025), students actively complain about reduced control over final outcomes and difficulties editing AI-generated artefacts; and in Da S. Soares et al. (2019), where AI mediates the student interaction with learning materials and students cannot opt out of AI outputs during the learning activity.

**RQ4: Design Knowledge – Recommendations, Research Gaps and Future Research Paths**

Regarding design guidelines and recommendations about HAIC/HI extracted from the revised papers (IQ4.1), our inductive analysis revealed several recurring themes. One expected theme in a field centered on collaboration is the quest for a *balanced approach towards the roles and affordances of both human and AI actors* (7 papers). Many related recommendations take a decidedly humanistic perspective, *stressing that AI cannot and should not substitute the learner (or instructor)* (12 papers). Accordingly, multiple works reference the *need for awareness, autonomy, human oversight, and agency* (10 papers), calling for humans to monitor and intervene on AI outputs while preserving human leadership. *Symmetry in roles and interactions* may also reinforce "collaborative" relationships (Flavin et al., 2025) and find advantages in *AI agents assuming various roles/personas*, thus being perceived as realistic, competent, and dependable partners (Edwards et al., 2025).

The reviewed papers offer a wide variety of more concrete design guidelines. These include using *gamification* to increase motivation (Aslan et al., 2024), having *humans double-check AI outputs* ("we encourage graduate students to reassess their SMART goals and action plans with their human mentors for personalized support" Chang et al., 2024, p. 14) and *integrating interventions in wider contexts* (e.g., going beyond coding to computational modeling, in Chen et al., 2024). Other proposals include *combining AI agents with self-monitoring SRL (Self-Regulated Learning) checklists* to reduce machine dominance and increase humanization of feedback (Darvishi et al., 2024) and leveraging AI *misrecognition as a learning opportunity* rather than merely an error (Song et al., 2022). Since AI systems "are particularly good at recognizing errors or misunderstandings in students' tasks" (Weber et al., 2025, p. 669), instructional and technology designers can use them to *spot specific errors and provide targeted formative feedback* at scale. One study proposed minimizing interaction effort for learners with disabilities (Da S. Soares et al., 2019). Others recommend *multiple feedback sources* "comprising qualitative [peer feedback] from teachers, quantitative multimodal data analysis recognized by AI, and professional knowledge provided by experts" (J. Yu et al., 2025, p. 590). Y. Fan et al. (2025) caution about *short- vs. long-term trade-offs*: AI agents "can significantly improve short-term task performance, but they may not boost intrinsic motivation and knowledge gain and transfer" (p.490). Several authors also address *ethical concerns*: trust, transparency, and bias (8 papers). There is also an important strand of papers highlighting the *need for adaptivity and personalization* (12 papers), rather than reliance on general-purpose chatbots.

Beyond these general guidelines, many authors stress the *importance of the interaction process and workflow* (17 papers), including structuring and sequencing interactions (particularly, in inquiry-based, project-based, or problem-based learning) and attending to sequential scaffolding and the end-to-end learning process. A further set of works (10 papers) underlines the need to *define specific roles or personas for AI agents* that fit learning process needs and promote critical thinking (e.g., Urban et al., 2024).

Turning to the human side of HAIC/HI, several works identify *desired properties of human learners* in (human-AI) collaborative contexts. A number of them emphasize developing *critical thinking capabilities* (12 papers) and *creativity* (4 papers), not only as 21st-century skills, but also as instruments for a *balanced partnership that avoids over-reliance on AI* (7 papers). *Metacognitive strategies* (8 papers) are similarly recommended, as metacognition and self- and social-regulation can both be triggered by human-AI interaction and serve as a buffer against imbalanced dependency. These learner-centric qualities are complemented by *engagement in reflection* (in- and on-action, cf. Schön, 2013) (10 papers) and using misrecognitions or failures as learning opportunities (Song et al., 2022). Authors also stress the *need for AI literacy*, including prompting skills (6 papers) and broader understanding of AI capabilities and limitations (12 papers). Other works call for *aligning HAIC/HI interventions with overall curriculum* (e.g., Hwang et al., 2025) *and learning objectives* (e.g., Patra et al., 2024) (9 papers), or for *creating a community of all related stakeholders*, e.g., educators, researchers, and developers (Alier et al., 2025).

Worth highlighting on its own is Bosch et al.'s (2025) work, noted above as a rare case of genuine "co-learning", in which both human and AI learn through interaction with each other and the environment. The authors provide recommendations for researchers aiming to *promote such co-learning*, including: carefully designing the AI's knowledge of task context, responsibilities, and others' roles; enabling in-action learning; encouraging learners to think critically and reflect on assumptions; making AI systems transparent through self-explanation and clarification-seeking; and ensuring symmetry of communication.

Regarding research gaps and the future research agenda (IQ4.2), 32 of the 62 reviewed papers call for *better understanding of HAIC/HI implications* for human learning (e.g., Bosch et al., 2025; Darvishi et al., 2024; Lee et al., 2025; Sengewald & Tremmel, 2024; Tang et al., 2024), particularly through longer-term studies (e.g., Quan et al., 2024; Rasnayaka et al., 2024). This is consistent with our analysis of the interventions in IQ1.8 and with broader AIED literature (e.g., Chiu, 2024; Weidlich et al., 2025). Eleven papers flag *negative risks* deserving further investigation: privacy and security (Adorni & Grosso, 2025), ethical impacts (Cai et al., 2024; Nguyen, Ilesanmi, et al., 2024; Rossato et al., 2024; Weber et al., 2025), trustworthiness (Cosentino et al., 2025; Figoli et al., 2022), over-reliance (Y. Fan et al., 2025), and cognitive overload (Kumar, Musabirov, et al., 2024). Notably, only 5 studies (Kim et al., 2024; Lee et al., 2025; Sankaranarayanan et al., 2020; Tang et al., 2024; E. Yu, 2023) explicitly identify *HAIC/HI structures and their associated interactions* as a primary target for future research, reinforcing the conclusion from IQ2.4 that the collaborative dimension of HAIC/HI remains peripheral to most reviewed works.

# Discussion

## Implications of the Review

Our analysis of 62 empirical studies on HAIC/HI for student learning reflects an *emerging technology-oriented field*: the prevalence of case studies, one-shot interventions, exploratory designs, and freeform AI use, all point to this state of flux. It thus remains premature to affirm the differential benefits of HAIC/HI approaches or particular collaboration structures for learning (cf. Yan, Greiff, et al., 2025 on the distinction between supporting learning and enhancing performance). Compounding this is a persistent definitional void: what counts as HAIC/HI remains unclear, and terms like "co-creation" or "teammate" are used loosely, even when AI acts as a mere tool to which tasks are offloaded. Whether "collaboration" is accurate or simply metaphorical – or should be abandoned altogether (see Sarkar, 2023) – remains debatable, given the typical disparity in human and AI goals and the fact that, in the overwhelming majority of cases, AIs do *not* learn (see Bosch et al., 2025 for an exception). "Hybrid intelligence" is therefore more literally precise, focusing less on the interaction and relationships between humans and AIs, and more on the overall phenomenon of composing different forms of intelligence.

Beyond the gaps identified in the reviewed papers themselves (see IQ4.2), our own assessment of the field points to several priorities for a *future research agenda*:

1. *More detailed descriptions of human-AI interactions*: as established in (CS)CL research (e.g., Dillenbourg et al., 1996; Stahl, 2006), granular accounts of "collaboration" are essential for understanding learning phenomena in these new settings.
2. *Exploration of more complex collaboration structures*: the structures surfaced in this review are relatively primitive (see IQ3.1). Emerging theoretical and empirical work is starting to identify more sophisticated micro- and macro-level HAIC/HI structures (e.g., Maya, 2024; Prieto et al., 2023), but empirical testing and comparison (against each other or against freeform interaction) remains lacking.
3. *More longitudinal studies*: moving beyond one-shot interventions is necessary to understand HAIC/HI as a potentially distinct form of learning, to rule out "novelty effects", and to examine both its benefits and unwanted side-effects (such as "metacognitive laziness" in Y. Fan et al., 2025).
4. *More rigorous evaluation of learning gains and learner skills*, given that assessing the effects of interventions on learning (not merely artifact quality or task efficiency), despite being the ostensible goal of the field, remains comparatively rare (see IQ1.9 above; and Yan, Greiff, et al., 2025 for a broader discussion in the context of GenAI in education).

The next subsection elaborates further implications concerning HAIC/HI structures and the design knowledge (technological and pedagogical) that this review yields, which represents our central focus in this paper.

**Where Is the Structure? *Towards* HAIC/HI for Learning Design Knowledge**

We began this SLR hoping to synthesize empirically backed design knowledge in the vein of Alexander's design patterns (recurrent problems in HAIC/HI for learning and their core solutions), themselves organized into pattern languages (Alexander et al., 1977). However, the lack of repetition in elicited structures (beyond the simple ones noted in IQ3.1) and the relative scarcity of studies comparing interventions' effects on learning force us to conclude *it is too early* for such formalization of design knowledge in the area of HAIC/HI for learning.

The three inductively identified collaboration structures in IQ3.1 could serve as *pattern seeds*, and relate to structures found in broader (i.e., not for learning) HAIC/HI literature. For instance, Wiethof and Bittner's (2021) "Decision Support" pattern (where the human provides initial input and the AI responds with a recommendation or explanation) resembles our "humans create, AI assesses" structure. Their "Exploration" pattern, where "the human gives implicit or explicit feedback on the returned results," resembles our "AI creates/answers, humans refine" structure. Notably, their "Creation" pattern, in which human and AI learning processes occur simultaneously and inseparably, did not appear in any of our reviewed papers.

Gomez et al.'s (2025) more recent review on HAIC/HI for decision-making identifies seven "interaction patterns", all of which involve the AI proposing an artifact to the human with varying degrees of proactivity. Some patterns (e.g., AI-guided dialogic user engagement, user-guided interactive adjustment) allow human feedback to refine AI proposals, but none suggests AI providing feedback on something the human previously developed. The authors themselves acknowledge the scarcity of "continuous user interaction scenarios, where user input is sustained and can receive AI feedback at any given moment". Our "Human creates, AI assesses" structure, however, may be more natural in education (where feedback is central to learning) than it is in decision support contexts.

To move toward more formalized design knowledge about HAIC/HI for learning, we can also draw from non-empirical sources in education and educational technology. For reusable macro-structures, well-established *collaboration flow patterns* from human-to-human collaborative learning offer a starting point: Lee et al. (2025) and Gutierrez-Ferré et al. (2024), for example, adapt the Jigsaw and Pyramid patterns (respectively) by introducing AI agents into them. Project-based learning literature similarly provides pedagogically grounded phase sequences that can be adapted for HAIC/HI scenarios (e.g., Fischer & Lanquillon, 2024; Hou et al., 2025).

Beyond AI "modes of contribution" (IQ2.3, cf. Cui & Yasseri, 2024) – such as assistant, coach, or teammate – the rarely-encountered "peer/equal" role, in which AI mimics a fellow student with a comparable knowledge level, merits deeper exploration. Adjacent literature reviews identify overlapping recurrent *roles*: learning partner (Deng & Yu, 2023), motivator, peer/co-learner (Han et al., 2025) – all of which can be repurposed in a way that is pedagogically sound. Sharples (2023) proposes more specific AI functions for learning conversations: possibility engine, Socratic opponent, collaboration coach… categories that partly overlap our empirically-sourced ones. Other emerging sources describe pedagogical roles for GenAI in adaptive simulations (Mollick et al., 2024), or offer reusable implementations of such roles (Mollick & Mollick, 2023), though formal evaluations of effectiveness remain scarce. At the *micro*-level (e.g., sentence openers, question prompts), LLMs' natural-language programmability means existing structures (e.g., social or epistemic scripts, see King, 2007; O’Donnell & Dansereau, 1992; Weinberger et al., 2005) can be directly incorporated into prompts in LLM-based HAIC/HI systems.

The paper's title question ("*where* is the structure?") also points to how structures are implemented. Our synthesis (IQ2.4, IQ3.1) reveals a majority of studies relying on socially-implemented structures, such as teacher instructions. This approach is flexible and low-cost but unreliable, depending on teacher capacity and orchestration load, which is often high (see Dillenbourg, 2013). Embedding structures into the technological interface (e.g., as UI steps or selectable prompts) is more reliable. LLMs introduce a "middle path": more flexible than hard-coded UI structures, though not guaranteed to produce the "intended interactions" (cf. Dillenbourg & Tchounikine, 2007), and potentially combinable with either of the other two implementation modes.

A related question is "*when* is the structure implemented?" Some works pre-set HAIC/HI structures rigidly at technology-design time; others, like Alier et al.'s (2025) LAMB framework, defer this to learning-design time; still others set the structure/constraints at enactment time through teacher instruction. Extending existing CSCL proposals, one could further hypothesize a "deploy-time" instantiation, where an abstract pedagogical design is customized for a particular classroom or student group (see Prieto et al., 2013).

More work is needed to refine these proto-patterns into a coherent pattern language that supports researchers, designers, and educators in designing HAIC/HI learning experiences.

Accordingly, more empirical research is also needed to quantify the relative effects (both intended and undesired) of different AI uses in learning (Y. Fan et al., 2025; Weidlich et al., 2025; Yan, Pammer-Schindler, et al., 2025).

## Limitations and Future Research

The present SLR is not without *limitations*. Like any SLR, our query may have missed relevant synonyms, though our initial exploratory search (see "Methodology") sought to minimize this. A further limitation is the explosive growth of GenAI research. Our emphasis on manual reading and processing, chosen to enable deeper understanding of the nuances of collaboration/interaction processes described in the papers (vs. wider, semi-automated approaches using LLM technology, with their advantages and disadvantages, Scherbakov et al., 2025; Tosi, 2025; Treviño & Arias-Carrión, 2025) made for a slower review and writeup. We therefore acknowledge that, by publication, some findings may already be outdated. That said, our unsystematic reading of more recent sources appears to confirm the general trends identified (definitional fuzziness, lack of interaction detail, failure to measure learning – see, e.g., Ong et al., 2026; F. Zhang et al., 2025). Regular updates to the review (as performed in 2025 following our initial 2024 wave) would be valuable as this dynamic field evolves. A further methodological limitation is our de-emphasizing of deductive codebook work and inter-coder agreement measures in the qualitative synthesis of the analyzed papers. While this reflects an interpretive stance on literature synthesis (where researchers, with their biases and backgrounds, serve as key research instruments, V. Braun & Clarke, 2006) it may reduce the perceived validity of our findings among more post-positivist researchers who favor quantitative measures of inter-subjectivity.

Another limitation of our findings stems from the comparatively thin descriptions of learner-AI interactions in many reviewed papers (see our research agenda above). This has constrained the richness of the HAIC/HI structures we could elicit from the empirical literature. As the research community grows to appreciate the importance of low-level interaction details for learning outcomes (as was discovered in collaborative learning research, Dillenbourg, 1999) richer interaction descriptions may become standard practice. This shift will also influence study designs, which currently rely heavily on post-hoc student surveys. Eventually, greater instrumentation of HAIC/HI environments and high-detail interaction analyses (common in fields like CSCL, see Hmelo-Silver et al., 2013) may follow.

Regarding *future research*, a priority should be convening groups of HAIC/HI (for learning) experts to engage in pattern synthesis, a practice well established in collaborative learning and learning design research (Baggetun et al., 2004; Goodyear, 2005) with the goal of deriving formal design patterns in the Alexandrian sense. Given the volume of recent work and likely tacit knowledge accumulating among researchers not yet codified in publications, formats such as Mor and colleagues’ "participatory pattern workshops" could be especially productive for building the design knowledge needed to support more effective HAIC/HI learning experiences (Mor et al., 2012; Mor & Winters, 2008).

## Conclusion

Our systematic literature review of empirical HAIC/HI works to support learning reveals a research area in flux and explosive growth. Synthesis efforts like this paper or the abovementioned pattern synthesis may help "settle the dust" on murky questions about whether having learners collaborate with generative AI chatbots, agents, and other elements, is beneficial (and for what, and to what extent) beyond the obvious answer (it depends). They may also help us understand the ethical implications of AI's use (e.g., the subsequent AI access gap, see Hammerschmidt). The current terminological fuzziness and definitional voids signal that several debates are overdue: whether "human-AI collaboration for learning" is even the right term (cf. Sarkar), and whether "AI learning" should also be among our goals, as fields like human-AI teaming have embraced (Van Den Bosch et al., 2019). These debates in turn reflect contrasting visions for this research area, and for educational technologies more broadly. A *humanist* view (still prevalent in most reviewed works, see IQ4.1 results) holds that human learning is the ultimate goal, learner agency is paramount, and AI, however agentic, remains a tool (e.g., Cukurova, 2025; Molenaar, 2022). A *post-humanist* view seems to be slowly emerging, perhaps signaled by terms like "hybrid intelligence," in which AI learning and agency matter in their own right, and a new form of intelligence (and, hence, learning), which is not entirely human (or goes beyond humanity) emerges. According to this emerging view, learning and collaboration are produced through entanglements of humans, technologies, materials, representations, and discourses (Sheridan et al., 2020). We look forward to participating in these debates and the interesting times (to quote the ancient Chinese curse) they may bring.

## Acknowledgements

The present work has been supported by grants PID2023-146692OB-C32, RYC2021-032273-I, and RYC2022-037806-I, all financed by MCIN/ AEI/ 10.13039/501100011033. These grants have been co-funded by the European Union's ERDF, ESF+, and "NextGenerationEU/PRTR". It has also been supported by the Regional Government of Castile and Leon and FEDER, under project grant VA176P23.

## Declaration of interest statement

The authors report there are no competing interests to declare.